# Development of Powder-in-Tube Processed Iron Pnictide Wires and Tapes

Yanwei Ma, Lei Wang, Yanpeng Qi, Zhaoshun Gao, Dongliang Wang, Xianping Zhang

*Abstract*—The development of PIT fabrication process of iron pnictide superconducting wires and tapes has been reviewed. Silver was found to be the best sheath material, since no reaction layer was observed between the silver sheath and the superconducting core. The grain connectivity of iron pnictide wires and tapes has been markedly improved by employing Ag or Pb as dopants. At present, critical current densities in excess of 3750 A/cm$^2$ ($I_c$ = 37.5 A) at 4.2 K have been achieved in Ag-sheathed SrKFeAs wires prepared with the above techniques, which is the highest value obtained in iron-based wires and tapes so far. Moreover, Ag-sheathed Sm-1111 superconducting tapes were successfully prepared by PIT method at temperatures as low as ~900°C, instead of commonly used temperatures of ~1200°C. These results demonstrate the feasibility of producing superconducting pnictide composite wires, even grain boundary properties require much more attention.

## I. INTRODUCTION

THE recent discoveries of new high-$T_c$ superconducting compounds based on the LaFeAsO system and subsequent studies on related materials including SmFeAsO (1111 type) and BaKFeAs (122 type) systems with a $T_c$ of 55 K and 38 K, respectively, have caused great interest [1]-[3], due to their high transition temperatures and extremely high upper critical fields $B_{c2}$ over 200 T [4]. Such critical field properties suggest that iron pnictides have strong potential for high-field applications above 30 K, where conventional superconductors cannot play a role owing to their low $T_c$'s.

For technological applications, such as magnets and cables, superconductors in the form of tapes or wires have to be developed. In 2008, our group produced the first LaFeAsO$_{0.9}$F$_{0.1}$ and SmFeAsO$_{1-x}$F$_x$ superconducting wires using Ta, or Nb tubes by the so-called powder-in-tube (PIT) method and they exhibited magnetic $J_c$ values of ~10$^3$ A/cm$^2$ at 5 K [5]-[6]. Soon after, SrKFeAs and FeSe wires were also fabricated by the IEE and NIMS, respectively [7]-[8]. The typical process of making 1111 type iron pnictide wires is as follows: $RE$ ($RE$: rare earth), As, $RE$F$_3$, Fe and Fe$_2$O$_3$ powders with stoichiometric composition are packed into metal tubes. The powder/metal composite tubes are cold worked into tapes and wires, and these tapes and wires are then heat treated at 1200°C for 40 hours.

Transport critical current density in high fields is very important in the view of potential applications. Unfortunately, we did not observe significant transport critical currents in the above wire samples [9], indicating that there is a significant depression of the superconducting order parameter at grain boundaries [10]-[11] or a reactive problem between the superconducting core and the sheath material for the wires obtained. In this paper, recent progress in research and development of iron-based wires and tapes with the enhanced transport properties through solving some of the above-mentioned issues is reviewed and the prospect for the next step is addressed.

## II. SILVER SHEATH MATERIAL

As we know, the PIT processed pnictide wires and tapes are usually subject to a sintering process at high temperatures in Argon atmosphere for tens of hours, therefore, we have to use sheath materials that are not reactive with pnictides. In our previous study [9], [12], Nb, Ta and Fe have been employed as sheath materials for wire fabrication, showing some reaction with pnictide during the final heat treatment. However, the absence of significant transport currents in wire samples has raised the concern that the reaction layer may have strong effect on the transport property.

Figure 1a and 1b show the optical images for typical transverse and longitudinal cross-sections of SmFeAsO$_{1-x}$F$_x$/Ta and Sr$_{0.6}$K$_{0.4}$Fe$_2$As$_2$/Nb wires after heat treatment, respectively. It is clear that for both kinds of wires, a reaction layer with a thickness of 10–30 μm was seen between the superconducting core and the tube sheath, although the thickness of reaction layer in SrKFeAs is much smaller than that in SmFeAsO$_{1-x}$F$_x$ wires due to the lower sintering temperature. These results demonstrate that the reaction between the core and the Nb or Ta sheath is quite harmful to the flow of transport critical current in the wire samples.

Fortunately, we recently found that compared to Fe, Ta and Nb tubes, Ag was the best sheath material for the fabrication of pnictide wires and tapes [13]. The transverse cross-sections of a typical Sr$_{0.6}$K$_{0.4}$Fe$_2$As$_2$/Ag/Fe wire and tape taken after heat treatments were shown in Fig. 1c. No reaction layer was

Manuscript received August 01 2010; accepted September 13, 2010. This work is partially supported by the Beijing Municipal Science and Technology Commission under Grant No. Z09010300820907 and National '973' Program (Grant No. 2011CBA00105).

Yanwei Ma, Lei Wang, Yanpeng Qi, Zhaoshun Gao, Dongliang Wang, Xianping Zhang are with the Key Laboratory of Applied Superconductivity, Institute of Electrical Engineering, Chinese Academy of Sciences, P. O. Box 2703, Beijing 100190, China (phone: 0086-10-82547129; fax: 0086-10-82547137; e-mail: ywma@mail.iee.ac.cn).

observed between the silver sheath and the superconducting core, indicating silver is benign in proximity to the compound at high temperatures. EDX line-scan (Fig.1d) further confirmed no diffusion of As or Sr into the volumes of Ag, which benefits superconducting properties of the $Sr_{0.6}K_{0.4}Fe_2As_2$ core. Most importantly, we succeeded in eliminating reaction layer by using silver as a sheath material, and all the wire and tape samples have shown the ability to transport superconducting current, typically around several Amperes [13].

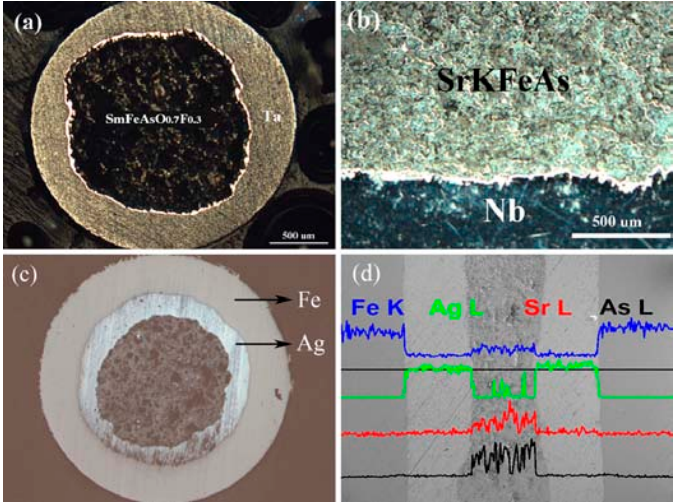

Fig. 1. The pnictide wires and tapes were fabricated by the PIT method. (a) The typical transverse cross-section of $SmFeAsO_{1-x}F_x$/Ta wire heated at 1180°C and (b) Longitudinal cross-section of $Sr_{0.6}K_{0.4}Fe_2As_2$/Nb wire heated at 850°C. (c) Transverse cross-sections of the $Sr_{0.6}K_{0.4}Fe_2As_2$/Ag/Fe wire and (d) EDX line-scan along the direction perpendicular to the longitude of Ag doped $Sr_{0.6}K_{0.4}Fe_2As_2$/Ag/Fe tape taken after heat-treatment.

### III. AG OR PB ADDITION TO IMPROVE THE GRAIN CONNECTIVITY

Poor transport properties in the pnictide wires are the principal limitation to technological applications. Recent studies on pnictide polycrystalline samples indicate that there are many liquid phases surrounding grains, resulting in a bad grain connection [14]. Hence, how to reduce the amount of grain boundary-wetting phase, as well as the crack density within grains, seems a key issue to be resolved.

Recently, we have found that the critical current density $J_c$ of polycrystalline $Sr_{0.6}K_{0.4}Fe_2As_2$ can be improved upon silver or lead addition [15]-[16]. In fact, we also prepared doped and undoped $Sr_{0.6}K_{0.4}Fe_2As_2$ bulks with adding an extra 10% K. As is evident from the figure 2, again, magnetic $J_c$ in the entire field region can be remarkably increased by Ag or Pb addition, compared to the pure sample. This positive effect of doping was further confirmed by the remanent magnetization measurements [17]. TEM investigation showed some Ag or Pb particles dispersed in the parent compound, with some residing between grains (see Fig. 3).

In order to understand the Ag effects, the microstructure of pure and Ag-doped samples was studied by HRTEM. It shows that with Ag addition, the glassy phases around individual grains and amorphous layer were effectively removed [15], and the connectivity among grains seemed to be improved by Ag doping, hence the $J_c$ enhancement.

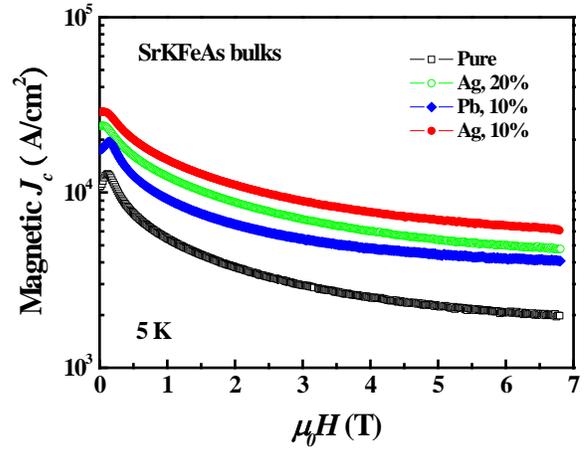

Fig. 2. Magnetic $J_c$ at 5 K as a function of applied field for pure, Ag- and Pb-added $Sr_{0.6}K_{0.4}Fe_2As_2$ bulk samples.

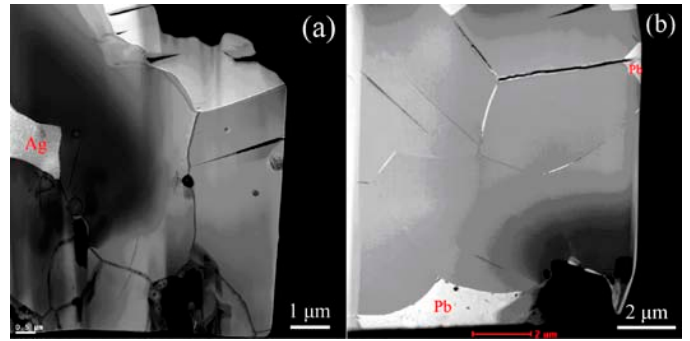

Fig. 3. TEM Images of (a) Ag- and (b) Pb-added $Sr_{0.6}K_{0.4}Fe_2As_2$ samples showing Ag or Pb particle.

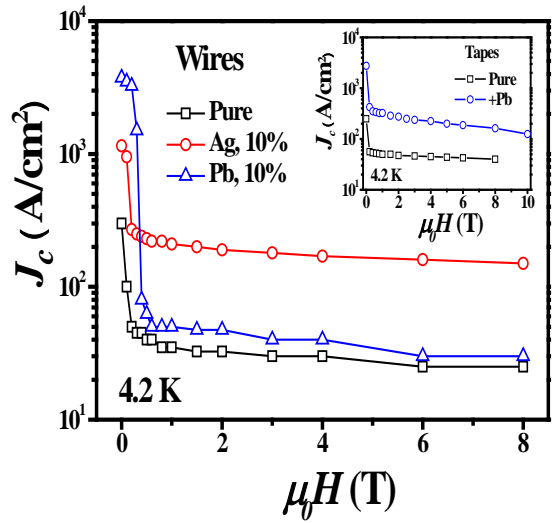

Fig. 4. Transport $J_c$-B properties at 4.2 K for Ag-sheathed pure, Ag- and Pb-doped $Sr_{0.6}K_{0.4}Fe_2As_2$ wires. The inset shows transport $J_c$-B curves of pure and Pb-doped $Sr_{0.6}K_{0.5}Fe_2As_2$ tapes.

As mentioned above, Ag sheath can successfully prevent from the reaction, and Ag or Pb can effectively improve the connectivity of grains. By using these two techniques, we succeeded in achieving significant transport critical currents in Fe/Ag/$Sr_{0.6}K_{0.4\sim0.5}Fe_2As_2$ composite wires and tapes [13], [18]. As shown in Fig. 4, the critical current density was strongly improved by Ag doping, increasing from 300 A/cm$^2$ for the pure wires to 1150 A/cm$^2$ for the doped wires in zero field. More importantly, a super current density of 100 A/cm$^2$ still flowed in the wire even under high fields up to 8 T. In particular, it is remarkable that the $J_c$ reached values as high as 3750 A/cm$^2$ in zero field for the Pb-doped wire, a factor of ~12 higher than that of the pure wires. The $J_c$ enhancement may be due to the elimination of cracks and enhanced connectivity. In addition, the transport $J_c$-B property in the higher K composition $Sr_{0.6}K_{0.5}Fe_2As_2$ tapes was further improved upon Pb doping, as shown in the inset of Fig.4. Therefore, the above data clearly indicate that the Ag or Pb addition is responsible for improving the superconducting properties of pnictide wires and tapes.

## IV. LOW TEMPERATURE FABRICATION OF SM-1111 WIRES AND TAPES

Generally, the SmFeAsO oxypnictides were synthesized at temperatures as high as 1150–1250°C, but this synthesis process is non-ideal due to the volatility of some components (F and As). On the other hand, low temperature fabrication process is required from the low cost point of view. Recently, we studied the effect of sintering temperature on the superconductivity of polycrystalline SmFeAsO$_{0.8}$F$_{0.2}$ bulks [19], and found that samples sintered at a low temperature clearly show high $T_c$, for example the $T_c$ of the samples sintered at 850°C is even above 50 K. Furthermore, the samples sintered at 900-1000°C show the higher RRR and the lower ρ (57 K), indicating the low impurity scattering and enhanced carrier density. This result suggests that annealing at a temperature of below 1000°C seems also suitable for obtaining high quality 1111 phase oxypnictides, compared to the commonly used temperatures of around 1200°C.

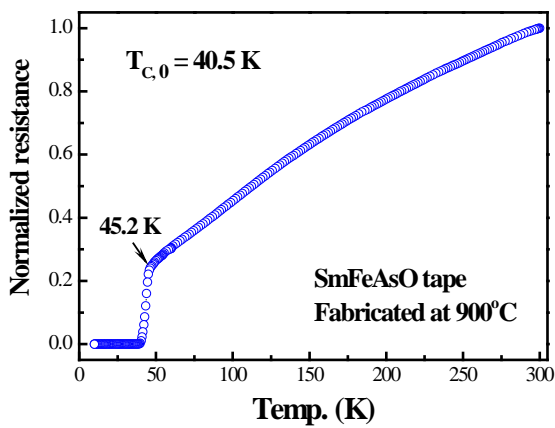

Fig. 5. Temperature dependence of resistance for the SmFeAsO$_{0.8}$F$_{0.2}$ tapes sintered at 900°C.

Our group then fabricated SmFeAsO (Sm-1111) superconducting wires by the PIT at a temperature as low as 900°C [20], about 300 K lower than those reported previously. In order to avoid the reaction problem, silver was used as a sheath material. Most interestingly, direct measurement of transport $J_c$ on Sm-1111 superconducting wires was obtained for the first time.

The XRD result shows that Sm-1111 tape samples consist of a main phase, Sm-1111, with a small amount of impurity phases present, ensuring the Sm-1111 phase can be formed at 900°C [20]. For SmFeAsO$_{0.8}$F$_{0.2}$ tapes, the onset $T_c$ occurs at 45.2 K, and zero resistance is achieved at 40.5 K, as shown in Fig. 5. Fig. 6 presents the transport critical current density $J_c$ as a function of field for SmFeAsO$_{0.8}$F$_{0.2}$ tapes synthesized at 900°C. Clearly, a transport $J_c$ as high as ~2700 A/cm$^2$ at 4.2 K and self-field has been observed. In the high field region, the $J_c$ is almost field independent, constant at ~180 A/cm$^2$. Although the transport critical current density $J_c$ in the superconducting wires is limited by the weak links between grains, higher $J_c$ values could be expected in textured materials, which is the on-going research.

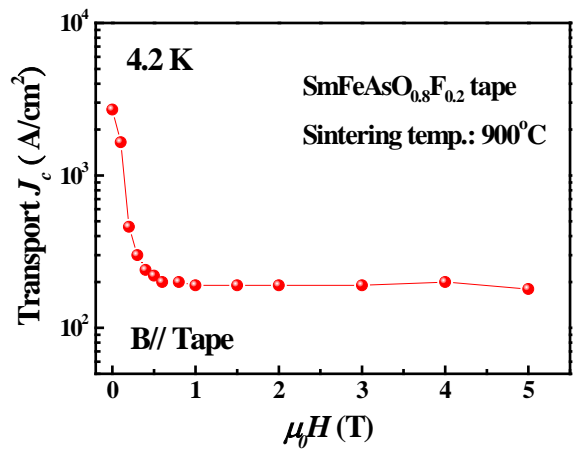

Fig. 6. Transport critical current density $J_c$ at 4.2 K as a function of field for SmFeAsO$_{0.8}$F$_{0.2}$ tapes heat-treated at 900°C.

## V. CONCLUSIONS

Through the investigation for the development of the PIT process for pnictide wires and tapes, the following several progresses were realized: Ag was found to be the best sheath material preventing from the reaction problem. The connectivity of grains was improved upon doping (Ag or Pb). On optimizing the effects of heat treatment, 900°C seems the suitable sintering temperature to achieve high performance doped pnictide superconductors [21]. Consequently, the $I_c$ value of 37.5 A was achieved in the 122 type wires with $J_c$ values of high than 3750 A/cm$^2$ at 4.2 K.

In addition, Ag-sheathed Sm-1111 superconducting wires were successfully prepared by PIT method at temperatures as low as ~900°C, instead of commonly used temperatures of 1200°C. Direct measurement of transport $J_c$ on Sm-1111 (or others RE-1111) superconducting wires was observed for the first time.

As mentioned above, the several important issues on the development of new pnictide wire process have independently been reviewed. In the next stage, further improvement in the $J_c$ properties is expected upon either optimization of processing parameters or achieving high grain alignment in analogy to high $T_c$ Bi-based cuprates.

ACKNOWLEDGMENT

We thank XiXiang Zhang at the KAUST for the TEM investigation, and S. Awaji and K. Watanabe at the IMR, Tohoku Univ. for the high field transport measurements. We are also indebted to H. H. Wen, D. C. Larbalestier, R. Flukiger, H. Kumakura, S. X. Dou and T. Matsushita for useful comments.